\documentclass[12pt,preprint]{aastex}
\newcommand{\chandra}{{\itshape Chandra}}
\newcommand{\asca}{{\itshape ASCA}}
\newcommand{\xmm}{{\itshape XMM-Newton}}

\newcommand{\rosat}{{\itshape ROSAT}}
\newcommand{\lum}{erg s$^{-1}$}
\newcommand{\flux}{erg cm$^{-2}$ s$^{-1}$}
\newcommand{\enedens}{erg cm$^{-3}$}
\newcommand{\colden}{cm$^{-2}$}
\newcommand{\ue}{$u_{\rm e}$}
\newcommand{\um}{$u_{\rm m}$}
\slugcomment{}
\shorttitle{Inverse Compton X-rays from  3C~98}
\shortauthors{N. Isobe et al.}
\begin{document}
\title{ 
The \xmm~Detection of Diffuse Inverse Compton X-rays
from Lobes of the FR-II Radio Galaxy 3C~98 
}
\author{N. Isobe\altaffilmark{1,2},
        K. Makishima\altaffilmark{1,3}, 
        M. Tashiro\altaffilmark{4}, and 
        S. Hong\altaffilmark{1,4} } 
\email{isobe@crab.riken.jp}
\altaffiltext{1}{Cosmic Radiation Laboratory, 
        the Institute of Physical and Chemical Research,
        2-1 Hirosawa, Wako, Saitama, 351-0198, Japan}
\altaffiltext{2}{ISS Science Project Office, 
        Institute of Space and Astronautical Science, 
        Japan Aerospace Exploration Agency (JAXA),
        2-1-1, Sengen, Tsukuba, Ibaraki, 305-8505, Japan.}
\altaffiltext{3}{Department of Physics, University of Tokyo,
        7-3-1, Hongo, Bunkyo, 113-0033, Japan.}
\altaffiltext{4}{Department of Physics, Saitama University,
        Shimo-Okubo, Saitama, 338-8570, Japan.}
\begin{abstract}
The \xmm~observation  
of the nearby FR-II radio galaxy 3C 98 is reported. 
In two exposures on the target, 
faint diffuse X-ray emission associated with the radio lobes 
was significantly detected,  
together with a bright X-ray active nucleus,
of which the 2 -- 10 keV intrinsic luminosity is 
$(4$ -- $8) \times 10^{42}$ \lum.
The EPIC spectra of the northern and southern lobes are reproduced 
by a single power law model modified by the Galactic absorption,
with a  photon index of  
$2.2_{~-0.5}^{~+0.6}$ and $1.7_{~-0.6}^{~+0.7}$ respectively.
These indices are consistent with that of the radio synchrotron spectrum, 
$1.73 \pm 0.01$. 
The luminosity of the northern and southern lobes are measured to be 
$ 8.3_{~-2.6}^{~+3.3} \times 10^{40}$ \lum~and 
$ 9.2_{~-4.3}^{~+5.7}  \times 10^{40}$ \lum~, respectively,
in the 0.7 -- 7 keV range. 
The diffuse X-ray emission is interpreted as an inverse-Compton emission,
produced when the synchrotron-emitting energetic electrons 
in the lobes scatter off the cosmic microwave background photons. 
The magnetic field in the lobes is calculated to be about $1.7~\mu$G, 
which is about 2.5 times lower than the value estimated 
under the minimum energy condition.  
The energy density of the electrons is inferred 
to exceed that in the magnetic fields 
by a factor of $40$ -- $50$.
\end{abstract} 

\section{Introduction}
Since the pioneering discoveries with \asca~and \rosat~
(e.g. Feigelson et al. 1995; Kaneda et al. 1995; Tashiro et al. 1998),
increasing numbers of diffuse hard X-ray emission have 
recently been detected from lobes in radio galaxies and quasars
(e.g; Brunetti et al. 2001; Isobe et al. 2002; Hardcastle et al. 2002). 
These diffuse X-ray photons are thought to arise via
inverse Compton (IC) scattering by electrons in the lobes, 
which also radiate synchrotron radio emission.
In most of these sources, the seed photons of the IC scattering in the lobes 
are provided by the cosmic microwave background (CMB) 
radiation (Harris \& Grindlay 1979),
or infra-red (IR) photons from the active nucleus (Brunetti et al. 1999). 
By comparing the IC X-ray and synchrotron radio intensities,
we can independently deduce energy densities 
of the electrons and magnetic fields 
in the lobes, \ue~and \um~respectively,
on condition that the seed-photon energy density is known.
Therefore, these IC X-rays from radio lobes provide 
a valuable tool to directly measure the energetics in lobes,
and infer those in jets, indirectly. 

The observed intensities of the IC X-rays from the lobes
frequently infer that \ue~significantly dominate \um~(e.g. Isobe 2002).  
The possibilities of magnetic fields 
smaller than those estimated under the equipartition condition by some factors 
are also reported in the knots and hot spots of radio galaxies and quasars
(e.g., Hardcastle et al., 2004; Kataoka \& Stawaltz, 2005).  
However, the sample of the IC X-ray lobes still remains small, 
due mainly to the limited sensitivity of the X-ray instruments. 
Obviously, \xmm~has a great potential in this research subject
owing to its large effective area 
and its moderately good angular resolution.
Actually, a few results have already been reported
based on the \xmm~data (Grandi et al. 2003; Croston et al. 2004).

We have conducted an \xmm~observation 
of a lobe-dominant radio galaxy 3C 98, 
and successfully detected the diffuse X-ray emission 
associated with its lobes.
As reported in the present paper, 
we find a clear dominance of \ue~over \um, in the lobes of this object.
Throughout the present paper, we assume a Hubble constant 
of $H_{0}= 75$ km sec$^{-1}$ Mpc$^{-1}$ and 
a deceleration parameter of $q_{0} = 0.5$. 

\section{Observation}
Located at a redshift of $z = 0.0306$ (33.7 kpc/1 arcmin; Schmidt, 1965),
3C~98 is a radio galaxy with an elliptical host (Zirbel, 1996).
It has a moderately high integrated radio flux 
of 10.3 Jy at 1.4 GHz (Laing, \& Peacock, 1980), 
and shows a synchrotron spectral energy index 
of $\alpha_{\rm R} =0.78$ 
between 178 and 750 MHz (Laing, Riley, \& Longair, 1983). 
Its VLA image (e.g., Leahy et al., 1997)
reveals a classical  FR II (Fanaroff, \& Riley, 1974) 
double-lobe morphology, 
with a relatively flat surface brightness distribution
and a total angular extent of $\sim 5'\times2'$. 
It dose not lie in a rich cluster environment, 
and is hence relatively free from hot thermal X-ray emission.
Moreover, a sign of extended X-ray emission 
on a scale of tens of arcseconds was reported 
based on an {\it ROSAT} observation (Hardcastle \& Worral 1999).
All these facts make this object very suitable 
to our search for the lobe IC X-rays.

Our 30 ksec \xmm~exposure on 3C 98 was performed 
on 2002 September 7 (hereafter, Exp1).
The EPIC MOS and PN CCD cameras  
on-board \xmm~were both operated in the nominal full frame mode. 
Because some normal stars in the field of view  
are brighter than the optical magnitude of 10,  
we inevitably adopted the medium and thick 
optical blocking filters for MOS and PN, respectively. 
Although these filters significantly reduce the efficiency 
below 1 keV, 
they little affect our study on hard-spectrum IC emission. 
Due to an operational failure, however, 
only about 60\% of the requested exposure was completed with MOS1, 
and no MOS2 data were taken in the observation. 
Therefore, another 10 ksec exposure was approved 
and conducted on 2003 February 5 (Exp2),
about 5 month after Exp1.  
The log of these two exposures are summarized 
in Table \ref{table:log}. 
 
We reduced the data using version 5.4.1 of 
the Science Analysis System (SAS). 
All of the obtained data were reprocessed, 
base on the latest Current Calibration Files (CCF),
just after the data of Exp2 were delivered to us (2003 April).  
We selected only those events with {\bf PATTERN $\le$ 12} 
(single, double triple and quadruple events) for MOS,
and {\bf PATTERN $\le$ 4} (single and double events.) for PN.
In addition, we also imposed the criterion of 
{\bf \#XMMEA\_EM} and {\bf \#XMMEA\_EP} for MOS and PN respectively, 
in order to filter out artifact events. 
As is frequently reported, the MOS and the PN backgrounds 
were both found to be highly variable, 
and so-called background flares 
significantly contaminated the data in both exposures.
Accordingly, we have selected the data obtained 
only when the 0.2 -- 15 keV count rate
was lower than 2.5 and 20 counts per sec for MOS and PN respectively, 
when accumulated within a source free region
which is about 96\% of the field of view. 
As a results, the remained good time interval became slightly short 
as summarized in Table \ref{table:log}.   
In addition to the above criteria;
we adopt only those events with {\bf FLAG == 0},
in the analysis below;
this strictly rejects events next to the edges of the CCD chips,
and next to bad pixels and columns.

\section{Results}
\subsection{X-ray Image}
Figure \ref{fig:image} shows the 0.2--12~keV EPIC image around 3C~98.
The image is smoothed with a two dimensional Gaussian function 
of $\sigma=5''$, 
after all the MOS and PN data obtained in both exposures
are summed up.
The 4.86 GHz VLA image (Bridle; unpublished) is superposed with contours. 
The bright X-ray source located near the center of the image
coincides with the optical host galaxy,
and also with the radio nucleus within $1''.5$, 
which is reasonable for the astrometric accuracy of the EPIC.   
In addition, we notice faint diffuse X-ray emission 
associated with the radio lobes, especially the northern one.

In order to visualize more clearly the diffuse X-ray emission,
we subtracted the point spread function (PSF) 
corresponding to the central point-like source at the host galaxy, 
based on the CCF data base. 
For this purpose, we used only the MOS data, 
because gaps of the PN CCD chips intersect the lobes. 
The in-orbit calibration of the MOS PSF is shown in detail,
by Ghizzardi (2001), at least for an on-axis point source.
The central point source is not so bright that 
the shape of the PSF is not distorted by significant CCD event pile-ups. 
We adequately took into account 
the energy dependence of the PSF, 
referring to the spectral information of the source,
shown in \S \ref{sec:nucle}. 
Before the subtraction, 
the PSF image is normalized to the observed one, 
by using the event counts within a circle of 15 arcsec radius,
around the X-ray peak.

\begin{figure}[htbp]
\plotone{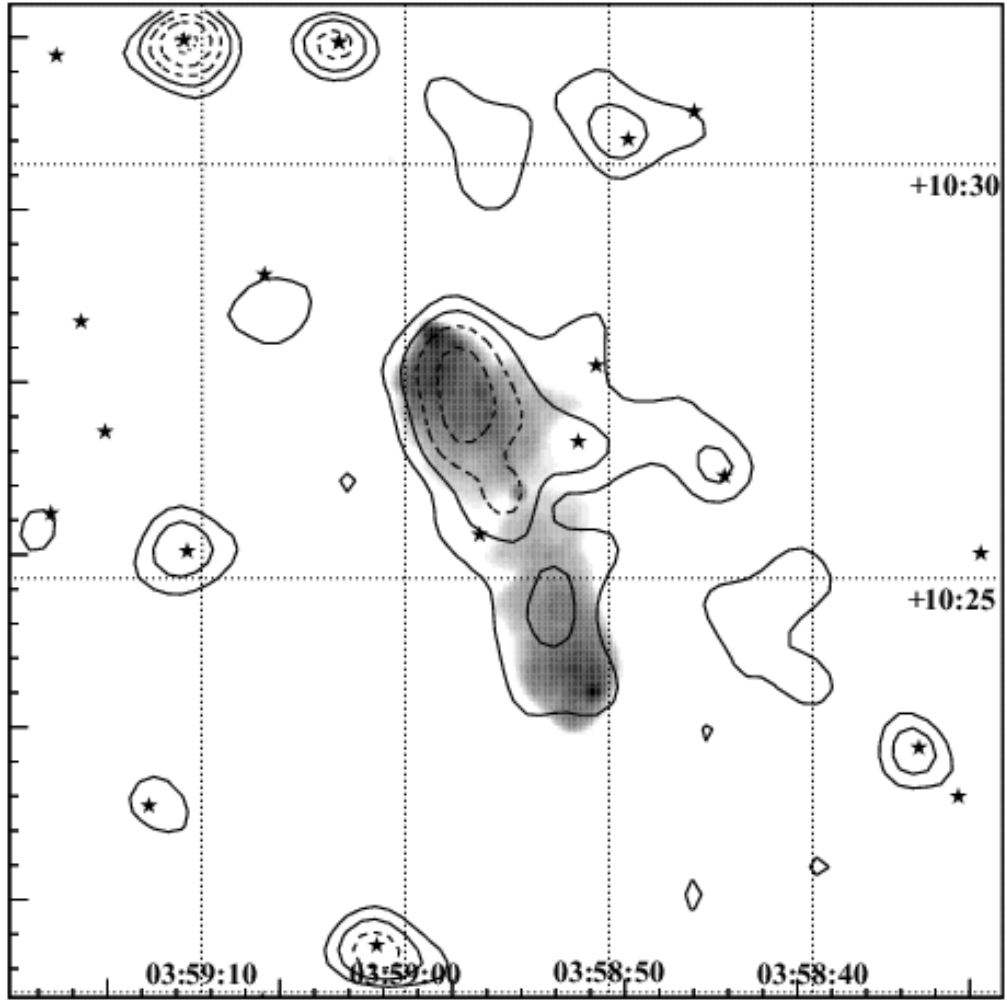}
\caption{
The 0.2 - 12 keV MOS contour image of 3C 98, 
which is superposed on the 4.86 GHz VLA gray scale image. 
The image is binned into $10''$ and  
heavily smoothed with a two dimensional Gaussian function 
of $\sigma = 20''$,
after subtracting the contributions from the host galaxy and its nucleus,
using the PSF image. 
Neither background subtraction nor exposure correction are performed.    
Six contours are written in a logarithmic scale,
between 2 and 5 counts per binned pixel.
The filled stars indicate the sources detected 
in the pipeline data processing, but excluding 3C 98 itself. 
}
\label{fig:subPSF}
\end{figure}

Figure \ref{fig:subPSF} shows the MOS image of 3C 98,
from which the PSF for the central source is subtracted,  
in 0.2 -- 12 keV.  
This image is heavily smoothed with a Gaussian function 
of $\sigma=20''$.
The diffuse X-ray emission is significantly detected,
in the figure, with some X-ray point sources.  
Moreover, 
its association with the lobes of 3C 98 is found to be clear,
and its overall angular extent almost coincides with that of the radio lobes.

We extracted linear X-ray brightness profiles 
parallel and orthogonal to the radio axis of 3C~98. 
in order to evaluate the significance of the diffuse X-ray emission. 
The integration strip for the parallel profile 
({\bf P$_{\rm 0}$} in Figure \ref{fig:image}) has an width of $2'.5$,
to contain all the radio structure of 3C~98.
The orthogonal profile utilizes a narrower strip,
{\bf O$_{\rm 0}$}, of a $1'.25$ width,
for minimizing contamination from the possible diffuse X-ray emission 
associated with the lobes.
We estimated the X-ray background for the parallel profile 
from the neighboring regions {\bf P$_{\rm b1}$} and {\bf P$_{\rm b2}$}, 
and that for the orthogonal profile  
from {\bf O$_{\rm b1}$} and {\bf O$_{\rm b2}$}.
All of these regions are shown in Figure \ref{fig:image} with dotted lines.  
 
The X-ray profiles thus obtained are shown in Figure \ref{fig:profile},
together with the radio ones accumulated within the same sky regions.
In the parallel profile,
an excess over the PSF is clearly 
found with a significance of $\gtrsim 4 \sigma$, 
accumulated between $0'$ and $3'$,
and $\sim 2 \sigma$ between $-3'$ and $0'$ from the X-ray peak.
On the other hand, 
the X-ray profile perpendicular to the lobes 
exhibits no significant excess over the PSF 
($\lesssim 1 \sigma$, between  $-3'$ and $3'$)
implying that the central source is point-like,
within the MOS angular resolution.
Therefore, we confirm that the EPIC image reveals   
the diffuse X-ray emission associated with the lobes of 3C 98.

\subsection{X-ray spectra of the host galaxy}
\label{sec:nucle}
\begin{figure}[htbp]
\plottwo{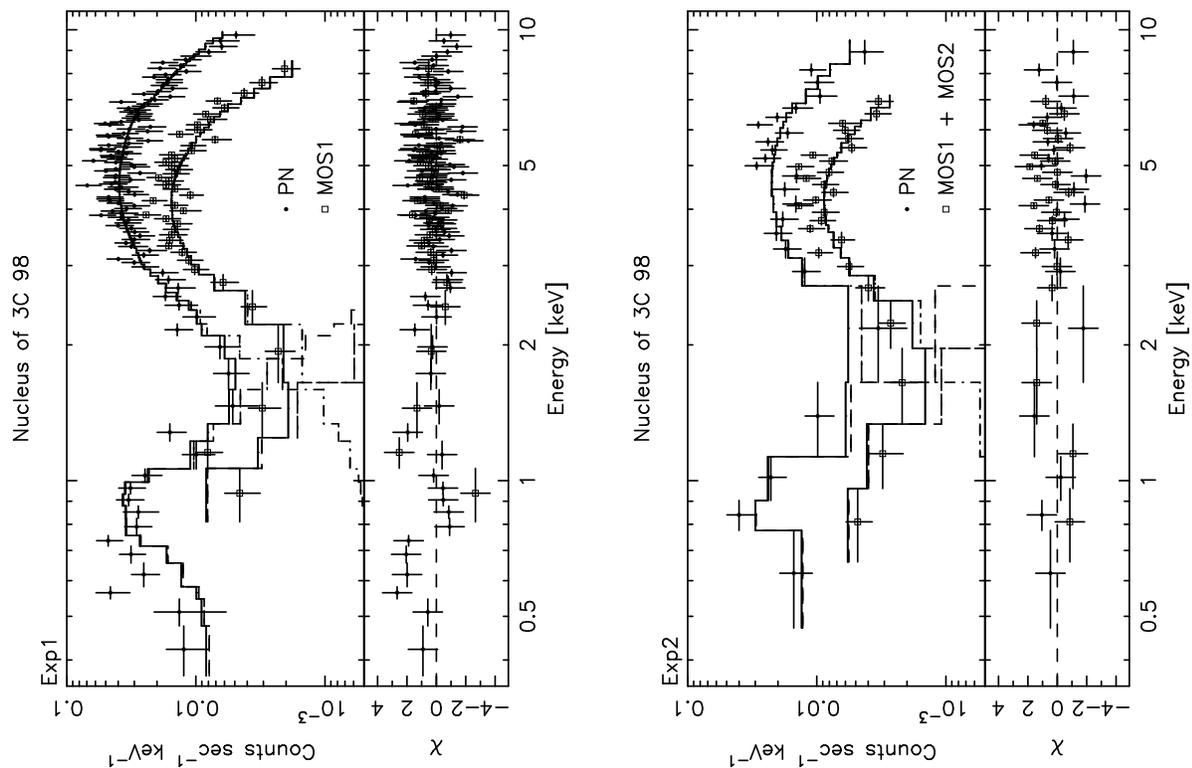}{f4b.ps}
\caption{The background-subtracted MOS and PN spectra 
of the host galaxy of 3C~98,
obtained in Exp 1 (left) and Exp 2 (right).
The best-fit model, consisting of an RS soft thin thermal emission 
and a heavily absorbed hard PL component, 
is shown with histograms.}
\label{fig:nucl_spec}
\end{figure}

We derived EPIC spectra within a circle 
of $0'.5$ ($16.9$ kpc) radius centered on the nucleus of 3C 98
(denoted as N in Figure \ref{fig:image}).
The background spectra were derived from
a neighboring source-free region with the same radius. 
We co-added the MOS1 and MOS2 data from Exp2. 
Figure \ref{fig:nucl_spec} shows the background-subtracted 
EPIC MOS and PN spectra of the host galaxy 
obtained in the individual exposures. 
They exhibit at least two spectral components; hard and soft ones. 
The hard component seems heavily absorbed, 
and its flux varied clearly between the two exposures. 
The soft and hard components are naturally attributed 
to the emission from hot gaseous halo and the active nucleus, 
respectively, of the host galaxy.  

We jointly fitted the MOS and PN spectra of each exposure
with a two component model, 
consisting of a soft Raymond-Smith (RS) thermal plasma component  
and an intrinsically absorbed hard power-law (PL) component.
Both components were subjected to the Galactic absorption,  
$N_{\rm H} = 1.17 \times 10^{21} $ 
\colden~(Hardcastle, \& Worrall, 1999, and reference therein). 
We fixed the abundance in the RS component at 0.4 times the solar value, 
which is typical among nearby elliptical galaxies (Matsushita et al., 2000).
The model has successfully  reproduced the data,
yielding the best-fit spectral parameters 
summarized in Table \ref{table:nucle}.
We found no clear evidence of Fe K line emission around 6 keV. 
An apparent data excess below $\sim 0.7$~keV, 
seen in the PN spectrum of Exp1, is statistically insignificant.  

Within the statistical uncertainty,
the spectral parameters of the soft RS component have stayed
constant between the two exposures.
The best-fit temperature and the 0.5 -- 10 keV luminosity
becomes $kT \sim 0.85$ keV 
and $L_{\rm RS} = (8 \sim 9) \times 10^{40}$ \lum, respectively,
in agreement with hot plasma emission from nearby elliptical galaxies 
(Matsushita et al., 2000).
The PL component has a flat spectral slope 
with a photon index of $\Gamma_{\rm X} \sim 1.4$,
and a high intrinsic absorption of $N_{\rm H} \sim 1 \times 10^{23}$ \colden;
these values did not change between the two exposures,
whereas its absorption-corrected luminosity 
was nearly halved meantime. 
These results support the identification of the hard component 
with the active nucleus emission. 
The intrinsic X-ray luminosity (Table \ref{table:nucle})
lies in the midst of those 
of radio galaxies (Sambruna et al. 1999).
The heavy absorption can be attributed to 
obscuration by outer regions of the accretion disk, 
in agreement with good symmetry between the two lobes.

\subsection{X-ray spectra of the lobes}
We accumulated  X-ray spectra of the northern and southern lobes 
within the circular regions LN and LS
shown in Figure \ref{fig:image}, respectively. 
The radius of LN is set to $1'.25$ (42.2 kpc),
and that of LS to  $1'.45$ (48.9 kpc),
to contain the whole lobe structure. 
To avoid contamination from the X-ray emission of the host galaxy,
we rejected events in a circular region of $1'$ (33.7 kpc) radius
centered on the nucleus, considering a typical size of a galactic halo. 
We summed the results over the two exposures, 
and co-added the MOS1 and MOS2 data. 

A background estimation is of crucial importance, 
especially for diffuse and faint X-ray sources. 
The background spectrum should be accumulated 
within the same CCD chip as LN and/or LS.
In the meantime, we had better select a region,
which is in a similar situation to the signal integration area,
with respect to the host galaxy.  
For PN, it is impossible to satisfy the second criterion, 
because of the gaps of the PN CCD chip. 
We, alternatively, select the circles of $1'.45$ radius (same as LS), 
BG1 and BG2 in Figure \ref{fig:image},
as the background region for LN and LS, respectively.
This is because a large fraction of LN and LS is 
within the same PN chip as BG1 and BG2.
A dotted circle within BG1 is removed, 
since an X-ray point source is detected in the position,
as shown in Figure \ref{fig:subPSF}. 
We also utilize the sum of the BG1 and BG2 spectra for MOS, 
after checking that the background around 3C 98 is spatially stable,
within the statistical uncertainties. 

\begin{figure}[htbp]
\plottwo{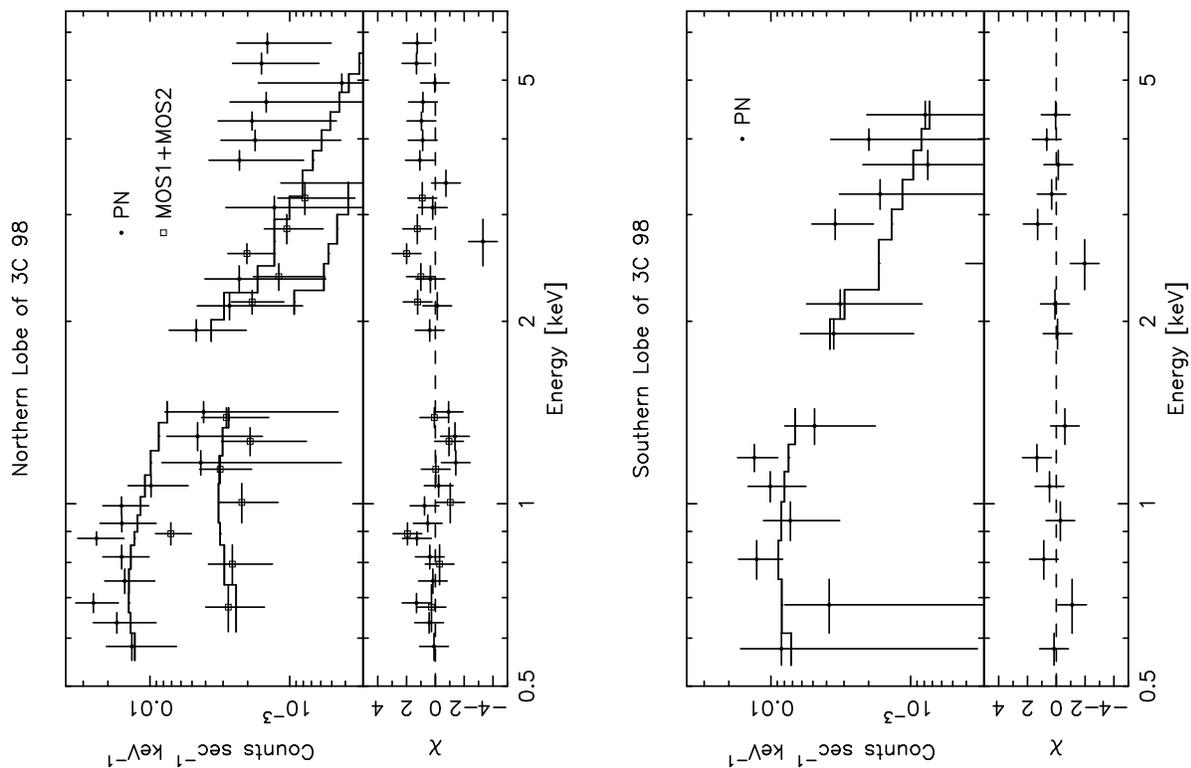}{f5b.ps}
\caption{The background-subtracted EPIC spectra of the northern (left) 
and the southern (right) lobes of 3C~98.
The data obtained in both exposure are summed up.
For the southern lobe, only the PN spectrum is shown and analyzed. 
Because of the prominent instrumental spectral feature, 
data in the 1.5 -- 1.7 keV range are discarded (see text).  
The best-fit model is shown with histograms. }
\label{fig:lobe_spec}
\end{figure}

The background-subtracted MOS and PN spectra 
of the northern and southern lobes of 3C 98 are shown  
in the left and right panels of Figure \ref{fig:lobe_spec}, respectively.
Because instrumental fluorescent lines severely contaminated 
the 1.5 -- 1.7~keV data,
we here and hereafter exclude this energy range from our spectral study.
The significance of the northern lobe signals is 
about $7.0 \sigma$ and $8.5 \sigma$
in the 0.6 -- 3.5~keV MOS and  0.5 -- 6~keV PN data, 
respectively, 
while that from the southern lobe signals is 
about $2 \sigma$ and $5.3 \sigma$
in the 0.6 -- 3.5~keV MOS and  0.5 -- 5~keV PN data. 
In the following,
we analyzed the spectra in these energy range, 
although we rejected the MOS data of the southern lobe, 
because of their slightly low signal significance.

We fitted the background-subtracted lobe spectra  
with a single PL model modified by the Galactic absorption.
This model has successfully reproduced the data
as shown with histograms in Figure \ref{fig:lobe_spec}, 
yielding the best fit spectral parameters 
summarized in Table \ref{table:lobe}. 
The best-fit model to the PN spectrum of the southern lobe 
is consistent with the MOS spectrum, within statistical errors.
Thus, the two lobes share almost the same set of parameters, 
though within rather large errors. 
In either lobe, 
the PL fit implies a 0.7 -- 7 keV luminosity 
of $\sim 9 \times 10^{40}$ \lum.

We also tried to describe the spectra 
with a thermal bremsstrahlung model (Bremss)
modified by the Galactic absorption.
As shown in Table \ref{table:lobe},
the fit turns out to be slightly worse, 
compared with the single PL model, especially for the northern lobe. 

\section{Discussion}
\label{sec:emission}
In addition to emission from the host galaxy,
including its nucleus, 
we have significantly detected diffuse faint X-ray emission 
from the lobes of 3C~98.
The X-ray spectra of the northern and southern lobes are 
reproduced by a single PL model of 
$\Gamma_{\rm X} = 2.2_{-0.5}^{+0.6}$ and $1.7_{-0.6}^{+0.7}$, respectively,
modified by Galactic absorption.

We are not able to exclude the thermal interpretation 
from the spectral fitting alone, 
and the best-fit Bremss model requires the thermal electron densities,
$n_{\rm th} \sim 2 \times 10^{-3}$ cm$^{-3}$, in the lobes of 3C 98.
However, numbers of studies on radio polarization effects indicate  
typically $n_{\rm th} \ll 10^{-3}$ cm$^{-3}$, 
within lobes of FR II radio galaxies 
(e.g., Burch 1979, Spangler \& Sakurai 1985). 
Therefore, we can conclude that 
thermal plasma in the lobes of 3C 98 themselves, 
has only a negligible contribution to the observed diffuse X-ray flux.   

\begin{figure}[htbp]
\plottwo{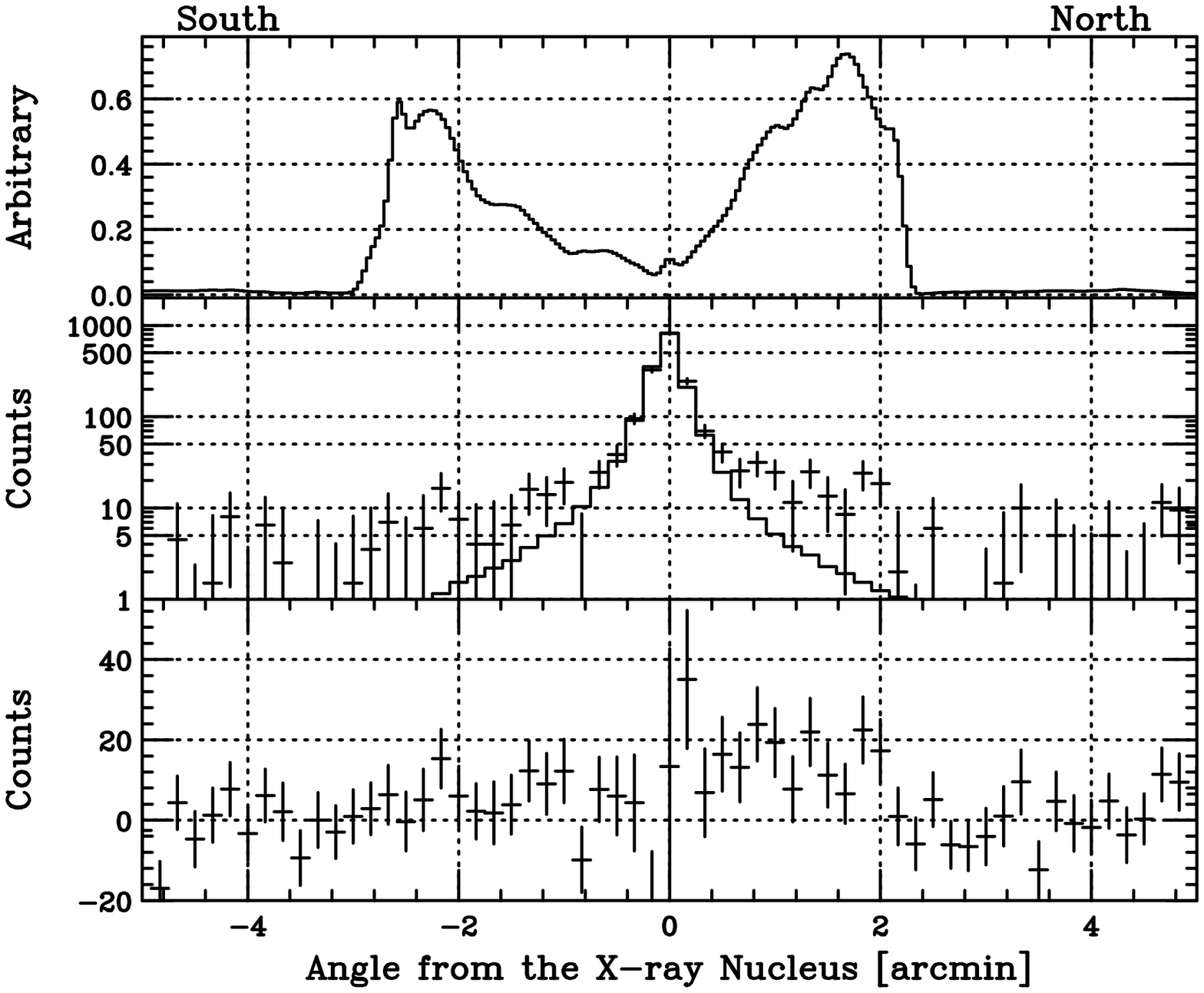}{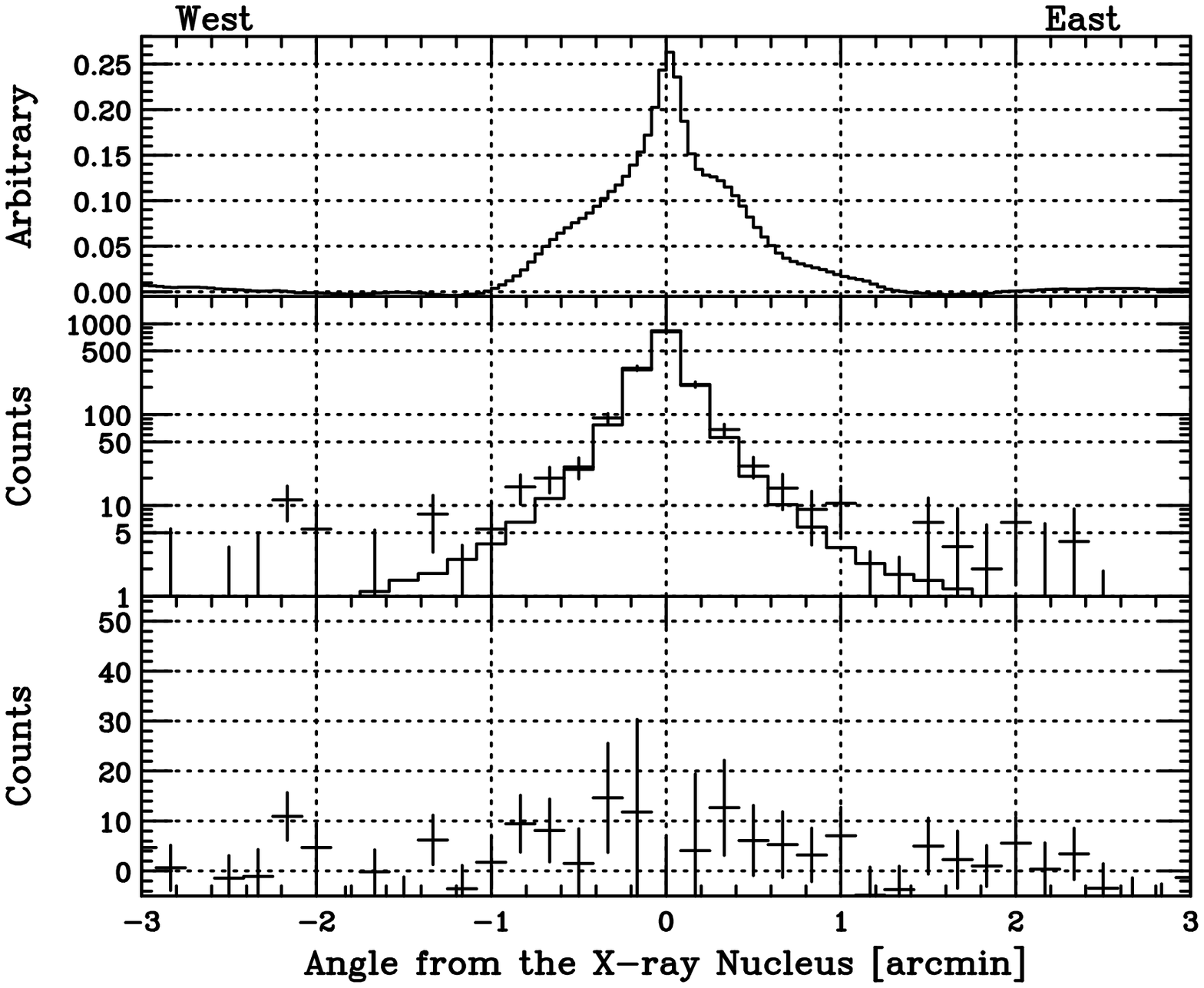}
\caption{The linear brightness profiles of 3C 98, 
parallel (left) and orthogonal (right) to the lobe axis, 
in the 4.86 GHz radio (top) and 
0.2 -- 12 keV X-ray (middle and bottom) bands. 
The X-ray profiles utilize only the MOS data (see text).
The middle panels show the background-subtracted X-ray profiles,
together with the MOS PSF ones (histograms), 
normalized by counts within a circle of 15 arcsec radius
around the X-ray peak.
The bottom panels also present the MOS profiles, 
but after subtracting the PSF ones.
The directions are indicated at the top of the figures.}
\label{fig:profile}
\end{figure}

Recent \chandra~and \xmm~observations frequently reveal that 
the lobes of radio galaxies produce the X-ray holes
in the thermal X-ray emission 
associated with their host galaxies and/or clusters of galaxies 
around them (e.g., McNamara et al, 2000; Finogurnov, \& Jones, 2001),
and as a result the holes are 
usually surrounded by the enhanced X-ray shell of 
the displaced thermal plasma.  
The best-fit Bremss temperature which is higher than 
that of the thermal plasma in the host galaxy (see \S \ref{sec:nucle}), 
may be consistent with the picture, 
in which the plasma around the lobes is heated 
through the bow shock (Kraft et al. 2003),
or by kinetic work (Croston et al. 2003) caused by its expansion.    
However, the thermal plasma to be displaced by the lobes 
is not detected at the distance corresponding to the lobes, 
in the direction perpendicular to the lobe axis 
(see the right panel of Figure \ref{fig:profile}).
The diffuse X-ray emission in 3C~98 
clearly has different spatial distribution; 
i.e., they seem to fill all over the lobes.
Moreover, 
the observed X-ray flux needs a thermal pressure 
($\gtrsim 10^{-11}$ dyne cm$^{-2}$) 
considerably higher than the non-thermal one of the lobes, 
and it should be difficult for the lobes to remove such a thermal plasma.  
Therefore, we strongly favor the non-thermal interpretation
for the diffuse X-ray emission.

\begin{figure}[htbp]
\plotone{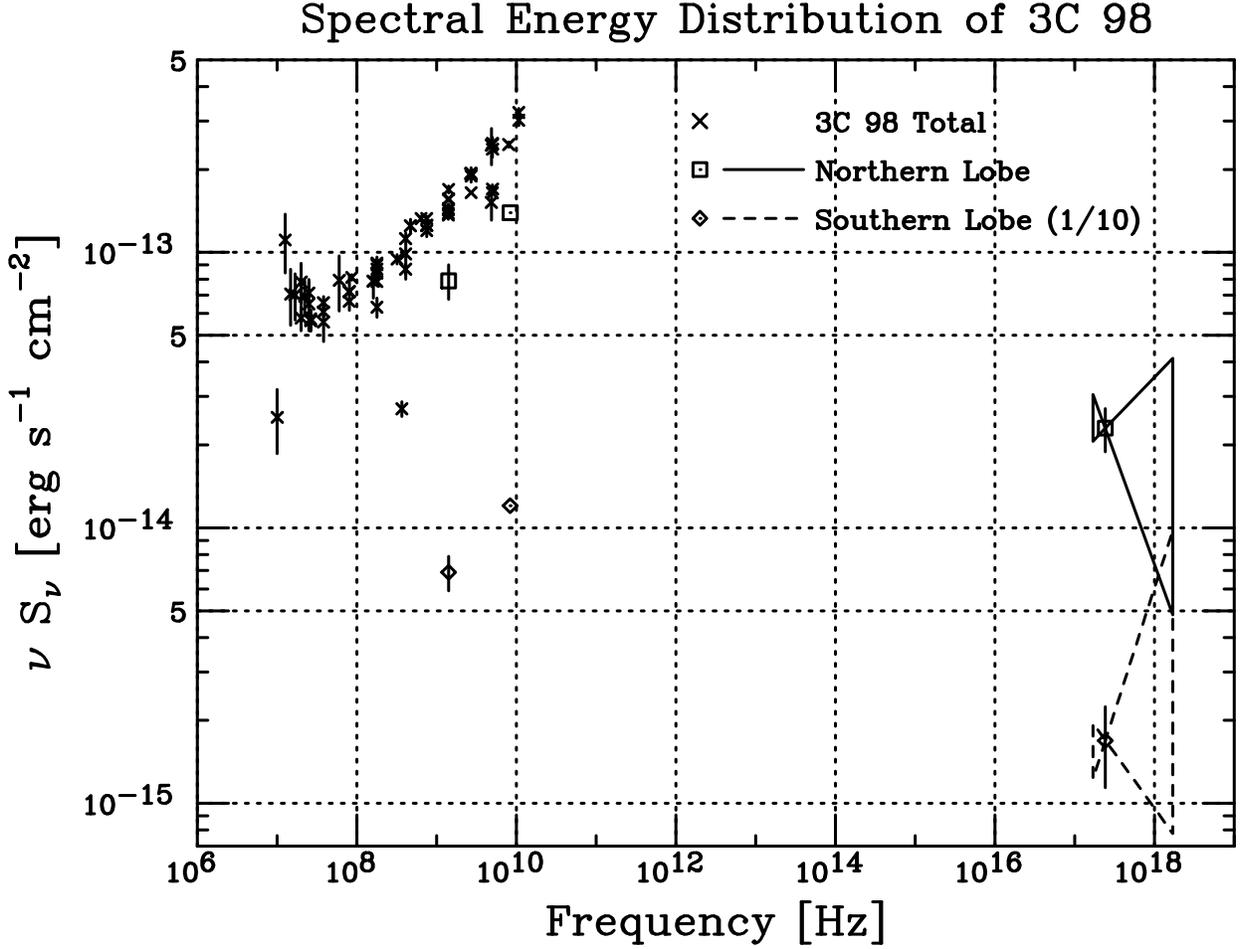}
\caption{The spectral energy distribution of the lobes of 3C 98.
The radio data, referring to the total flux densities,  
are collected from NASA/IPAC Extra galactic Database 
(NED; and references therein).
The data at 1.4 GHz and 8.4 GHz for individual lobes 
are taken from Mackay 1969 and Hardcastle et al. (1998), respectively. 
The solid and dotted ties show the PL model spectra  
to the northern and the southern lobe X-ray emission, respectively. 
Both radio and X-ray spectra of the southern lobe 
are shifted downward by an order of magnitude, 
to avoid overlap with the northern lobe spectra.
}
\label{fig:sed}
\end{figure}

Figure \ref{fig:sed} shows the radio and X-ray spectral energy distribution 
of the northern and southern lobes of 3C 98.
Radio data referring to the total flux of 3C 98 are also shown. 
The contributions from the components other than the lobes 
(such as the nucleus and hot spots) 
are reported to be very small (at most 5 \%;  Hardcastle et al. 1998).  
At least from 20 MHz and 10 GHz, 
the total radio synchrotron spectrum is well described by a single PL model 
(except for only a few deviating points); 
the flux density at 1.4 GHz and the photon index 
becomes $S_{\rm R} = 11.1 \pm 0.1 $ Jy 
and $\Gamma_{\rm R} = 1.73 \pm 0.01 $, respectively.
This value of $\Gamma_{\rm R}$ 
falls well within the range of two-point spectral index 
of both lobes between 1.4 GHz and 8.35 GHz, ($1.6 \sim 1.75$).
$\Gamma_{\rm R}$ is also consistent with the X-ray photon index of the lobes,  
within the statistical uncertainties. 
This agreement makes the IC X-ray interpretation fully self-consistent.  
Hence, we re-analyzed the X-ray spectra with the PL model 
of which the photon index is fixed at $\Gamma_{\rm R}$
and obtained the parameters shown in Table \ref{table:lobe}.

The non-thermal X-ray spectrum may alternatively be interpreted 
as a high-energy extension of the synchrotron radio spectrum.
However, as is clear from Figure \ref{fig:sed}, 
it is impossible to describe the radio and X-ray spectra simultaneously 
by synchrotron radiation from a single electron population  
with a PL energy distribution,
even if we invoke a synchrotron cutoff in the electron spectrum. 
Moreover, a synchrotron X-ray photon 
would need the electron Lorentz factor to exceed 
$ \gamma_{\rm e} \sim 4.5 \times 10^{8}~ 
               E_{\rm keV}^{~0.5} B_{ \mu {\rm G }}^{~-0.5} $
where $E_{\rm keV}$ is the energy of the synchrotron photon in keV, 
and $B_{ \mu {\rm G}}$ is the magnetic field strength in $\mu$G; 
in that case, the synchrotron cooling time scale of those electrons 
which are responsible for the synchrotron X-ray photons would   
become unrealistically short, less than about 
$ 5.5 \times 10^4
              E_{\rm keV}^{~-0.5} B_{ \mu {\rm G}}^{~-1.5}$ yr. 
Therefore, we conclude that our working hypothesis, 
namely the IC scattering of some seed photons by the synchrotron electron,
provides the most plausible explanation 
for the observed diffuse X-ray emission. 

The soft seed photons of the IC scattering in the lobes are
usually provided by either CMB photons or IR radiation from the nucleus 
(see \S1).
Base on the observed IR flux density, 
85~mJy at 25~$\mu$m (Golombek et al., 1988),
the IR luminosity of the nucleus of 3C 98 
is estimated to be $L_{\rm IR} \sim 2 \times 10^{43}$ \lum. 
Even though the orientation and the obscuration effects 
are taken into account (Heckman et al. 1994), 
$L_{\rm IR}$ dose not exceed $\sim 10^{44}$ \lum. 
This yields an IR photon energy density of 
$u_{\rm IR} \sim 10^{-14} (r/50~{\rm kpc})^{-2}$ \enedens,  
where $r$ is the distance from the nucleus in kpc. 
At $r \gtrsim 8$ kpc, 
$u_{\rm IR}$ becomes lower than the CMB energy density
$u_{\rm CMB} = 4.6 \times 10^{-13}$ \enedens~at the redshift of 3C~98.
We therefore conclude that $u_{\rm CMB}$ dominates over $u_{\rm IR}$
in the larger part of the lobes of 3C~98.  

\begin{figure}[htbp]
\plotone{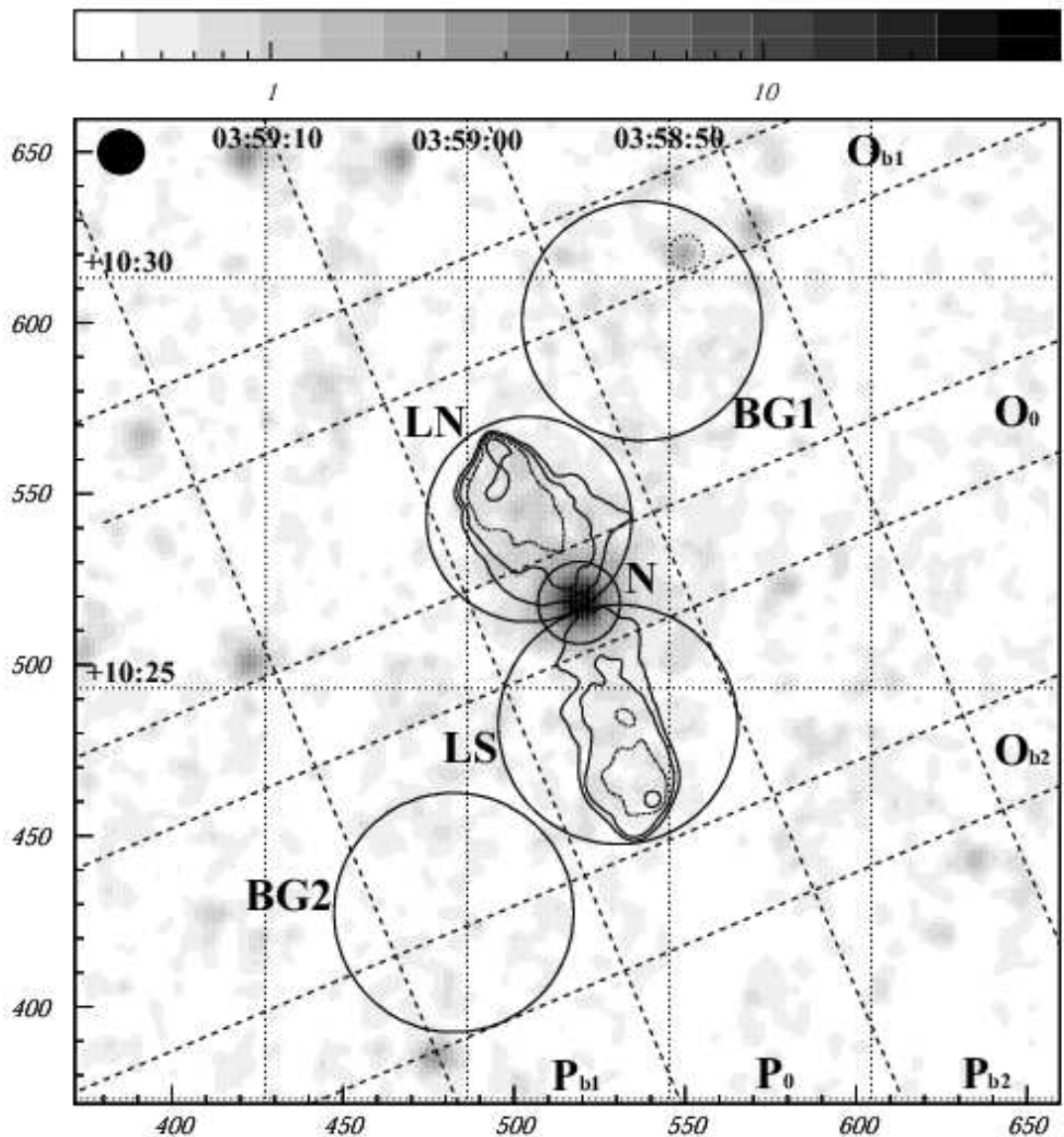}
\caption{The 0.2--12 keV EPIC MOS + PN image of 3C 98.
The image is binned into $2.5''$  and 
smoothed with a two dimension Gaussian function of $\sigma = 5''$.
Neither background subtraction nor exposure correction are performed.    
The 4.86 GHz VLA contours (Bridle, unpublished) are overlaid.
Solid circles indicate integration region 
for the signal (N, LN and LS) and background 
(BG1 with dotted circle rejected, and BG2) spectra.
The filled circle at the top-left shows the half energy width 
of the MOS and PN PSF (about 15'' radius).
Strips between dashed lines 
specify the regions used to obtain the signal (P$_0$ and O$_0$) 
and background (P$_{\rm b1}$, P$_{\rm b2}$, O$_{\rm b1}$, and O$_{\rm b2}$) 
linear brightness profiles.  
}
\label{fig:image}
\end{figure}

Table \ref{table:param} summarizes relevant parameters 
to diagnose energetics in the lobes.
We assumed the shape of each lobe to be a simple ellipsoid, 
with the major and minor axes evaluated
from the radio VLA image (Figure \ref{fig:image}),
to calculate the volume $V$ of the lobes. 
The synchrotron radio flux densities of the individual lobes 
were derived from the PL fit 
to the total radio spectrum of Figure \ref{fig:sed}, 
and their relative contributions taken from Hardcastle et al. (1998).
We calculated the corresponding IC X-ray flux density 
from the PL fit to the lobe X-ray spectra 
with the photon index fixed at $\Gamma_{\rm R} = 1.73$;
this specifies the index of the electron number density spectrum
as $ 2\Gamma_{\rm R} - 1 = 2.46$.
To evaluate \ue, 
we assume an electron filling factor of unity, 
and integrated the electron spectrum
between the Lorentz factor of $\gamma_{\rm e} = 10^3$ and $10^5$, 
because the synchrotron or IC radiation 
from such electrons is directly observable; 
the lower limit corresponds to 1 keV IC X-ray photons,
while the upper limit to 
synchrotron photons of $\sim 10~ B_{ \mu {\rm G}} $ GHz. 

Referring to Harris and Grindlay (1979),
we have determined \ue, \um, and the corresponding magnetic field $B$,
as show in Table \ref{table:param}.
The physical parameters are found to be quite similar 
between the two lobes. 
In both lobes of 3C~98, 
the electrons are inferred to highly dominate over the magnetic fields, 
parameterized as $u_{\rm e}/u_{\rm m} = (40 - 50)$.  
This means that the magnetic field in the lobes is 
about 2.5 times lower than $B_{\rm me}$, 
where $B_{\rm me}$ is magnetic field strength 
which is estimated  under the minimun energy condition 
without proton contribution (Miley 1980).

In order to examine the uncertainty of  the result 
shown in Table \ref{table:param}, 
we evaluate possible systematic uncertainties 
in $S_{\rm R}$, $S_{\rm X}$, $\Gamma_{\rm R}$ and $V$,
all of which are fundamental observable parameters 
for the calculation of \ue~and \um.
Based on the following discussion, 
the uncertainties in \ue~and \um
are estimated to be at most 25 \% and 35 \%, respectively,
which can reduce the electron dominance $u_{\rm e}/u_{\rm m}$,
only a factor of two.  
Therefore, our conclusion basically holds.  

We consider that 
the uncertainty in $S_{\rm R}$ dose not exceed 15 \%,
based on the radio spectrum in Figure \ref{fig:sed}.     
On the other hand, $S_{\rm X}$ is thought 
to have an error of about 20 \%,
mainly due to the integration regions for the signal and background spectrum.  
Obviously, these errors almost directly propagate to \ue~and/or \um,
since \ue~and \um~is in proportion to $S_{\rm X}$,
and to $ (S_{\rm R} S_{\rm X}^{-1})^{2/\Gamma}$, respectively.

If the major axis of the lobes (and hence, the jets) of 3C 98 
is not precisely perpendicular to our line of sight, 
the adopted volume of the lobes become smaller than the actual one.
This results in an overestimation of \ue~with no impact on \um,  
because of inverse proportionality of \ue~to $V$. 
Based on the theoretical prediction of jet to counter jet brightness ratio, 
taking a relativistic boosting into account (e.g, Giovannini, et al., 2001),
and on the observed flux densities of the northern and 
the southern inner jets of 3C 98 (Hardcastle et al. 1998),
we estimated the jet angle to our line of sight to be about 75 degree. 
The angle put only a negligible effect on $V$ and \ue~of the lobes
(at most 5 \%). 

The smaller spectral index can  artificially 
enhance the electron dominance.
Our employment of $\Gamma_{\rm R}$ as the spectral index of the lobes 
should be safely justified, 
because the sum flux densities of the lobes 
highly dominates the total flux density of 3C 98 (about 95 \%), 
and the spectral shape of the lobes are very similar to each other,
and also to that of the total spectrum.    
Base on the radio spectrum of Figure \ref{fig:sed} alone,
we cannot fully reject  
the possibility of the index larger than $\Gamma_{\rm R}$.
However, even the upper limit on the two-point index of the lobes, 1.75, 
will reduce \ue~only about 7 \%, 
and simultaneously enlarge \um~about 15 \%,
compared with the values in Table \ref{table:param}. 

\begin{figure}[htbp]
\plotone{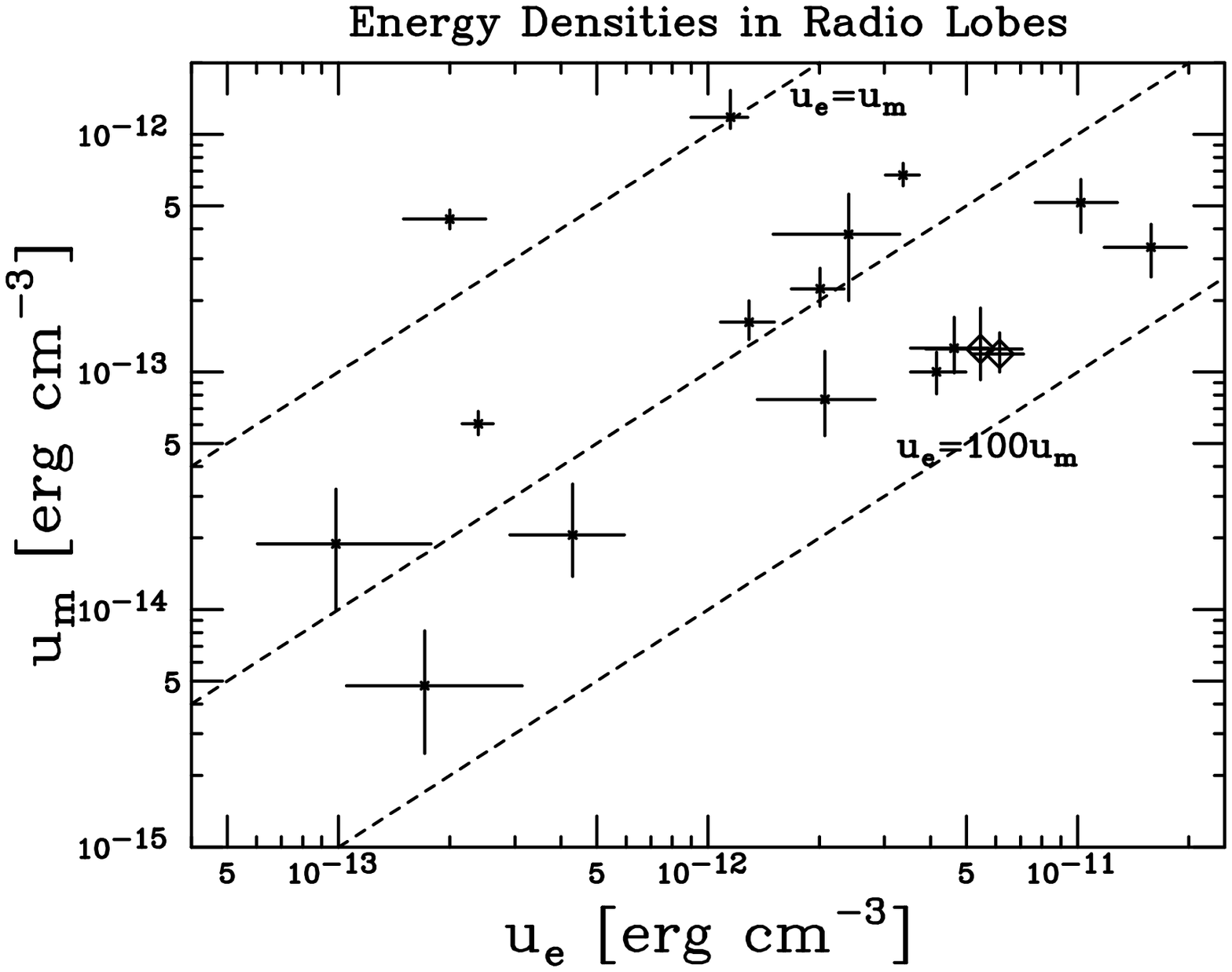}
\caption{The relation between $u_{\rm e}$ and $u_{\rm m}$  
in lobes of 3C 98, together with previous results on the other lobes  
(Tashiro et al. 1998; Brunetti et al. 1999;
Tashiro et al. 2001; Isobe 2002; Isobe et al. 2002; 
Comastri et al. 2003; Grandi et al. 2003; Croston et al. 2004).
The northern and southern lobes of 3C 98 are indicated 
by the right and left open diamonds, respectively. }
\label{fig:ueum}
\end{figure}

Figure \ref{fig:ueum} is a compilation of the reported IC X-ray detections 
in terms of \ue~and \um.
This figure suggest that 
the electron dominance of $u_{\rm e}/ u_{\rm m} \sim 10$ 
(corresponding to $B \lesssim 0.5~B_{\rm me}$, 
slightly depending on the spectral index),
is typically observed in the lobes of radio galaxies,
from which the IC X-rays are detected. 
Recently, Croston et al. (2005) have reported 
an almost consistent result of $B \sim 0.7~B_{\rm me}$ in lobes,
based on the \chandra~and \xmm~observations 
of 33 FR-II radio galaxies and quasars.  
The figure indicates that  
the present result on 3C~98 is very typical among these measurements. 

The observed electron dominance will raise an important question;
what does confine the non-thermal plasma within the lobes ?     
Although we regard that it is currently difficult to answer the question,
the spatial distribution of \ue~and \um~in the lobes 
may provide us an useful clue to the problem. 
In several lobes in which the electron dominance is already reported, 
almost uniform or center-filled distributions of IC X-rays are found,
in spite of a rim-brightening synchrotron radio feature 
(Tashiro et al. 1998; Tashiro et al. 2001; 
Isobe et al. 2002; Comastri et al. 2003).
A comparison of the X-ray and radio distributions indicates 
that the magnetic field is enhanced toward the edge of the lobes, 
to become even closer to the equipartition 
(Tashiro et al., 2001;  Isobe et al., 2002),
while the electrons are relatively uniformly distributed over the lobes.
As shown in Figure \ref{fig:profile}, 
the lobes of 3C 98 have the tendency 
though with a little large statistical errors. 
However, the clear solution to the issue inevitably needs 
more detailed observations with higher statistics, 
and theoretical considerations.

\acknowledgments
We are gratefull to the anonymous referee 
for his constructive comments to improve the present paper. 
We thank Dr. I. Takahashi for his guidance of the \xmm~data analysis,
especially the background estimation.
This paper is based on observations obtained with \xmm, 
an ESA science mission with instruments and contributions 
directly funded by ESA Member States and NASA.  
This research has made use of the NASA/IPAC Extra galactic Database
(NED) (the Jet Propulsion Laboratory, California Institute
of Technology, the National Aeronautics and Space Administration). 
The unpublished VLA image of 3C~98 (Bridle) was downloaded from 
``An Atlas of DRAGNs''  (http://www.jb.man.ac.uk/atlas), 
edited by Leahy, Bridle, \& Strom.      
   

\clearpage
\begin{table}[htbp]
\caption{Log of \xmm~observations of 3C 98.}
\label{table:log}
\begin{tabular}{llcc}
\hline\hline 
        &       &  Exp1          & Exp2    \\
\hline       
Date     &       &  2002.9.7      & 2003.2.5         \\  
Mode     &       &  \multicolumn{2}{c}{Full Frame}                 \\
Filter   & MOS   &  \multicolumn{2}{c}{Thick}                 \\
         & PN    &  \multicolumn{2}{c}{Medium}                 \\ 
GTI\tablenotemark{a} & MOS1  &  13.7   &  6.65       \\
         & MOS2  &   --    &  7.16       \\
         & PN    &  10.6   &  3.7       \\
\hline       
\end{tabular}
\tablenotetext{a}{Good time interval in ksec, 
which remains after rejecting high background periods.}
\end{table}

\clearpage
\begin{table}[htbp]
\caption{The best-fit spectral parameters of the nucleus of 3C~98}
\label{table:nucle}
\begin{tabular}{llcc}
\hline\hline 
               &                                & Exp1                    &  Exp2  \\
\hline       
\multicolumn{2}{l}{MOS Count Rate\tablenotemark{a}} &   $6.4 \pm 0.2 $    & $ 4.2 \pm 0.2$       \\
\multicolumn{2}{l}{PN Count Rate\tablenotemark{a}}  &   $21.7 \pm 0.5$    & $ 12.5 \pm 0.6 $         \\
\hline       
\multicolumn{2}{l}{Soft RS component}           &                         & \\
\hspace{0.5cm} & $kT$ (keV)                     & $0.85_{-0.05}^{+0.04}$  & $0.88_{-0.04}^{+0.07}$\\
\hspace{0.5cm} & $F_{\rm RS}$\tablenotemark{b}  & $4.8_{-0.8}^{+1.8}$     & $4.4_{-1.0}^{+1.6}$\\
\hspace{0.5cm} & $L_{\rm RS}$\tablenotemark{c}  & $8.8_{-1.4}^{+3.3}$     & $8.0_{-1.8}^{+3.0}$\\
\hline       
\multicolumn{2}{l}{Hard PL component}           &&\\
\hspace{0.5cm} & $\Gamma_{\rm X}$               & $1.43_{-0.19}^{+0.21}$  & $1.40_{-0.40}^{+0.41}$ \\
\hspace{0.5cm} & $N_{\rm H}$\tablenotemark{d}   & $1.08_{-0.12}^{+0.13}$  & $1.07_{-0.21}^{+0.26}$\\
\hspace{0.5cm} & $F_{\rm X}$\tablenotemark{e}   & $4.5_{-0.3}^{+0.4}$     & $2.6_{-0.3}^{+0.4}$\\
\hspace{0.5cm} & $L_{\rm X}$\tablenotemark{f}   & $8.0_{-0.5}^{+0.6}$     & $4.6_{-0.6}^{+0.7}$\\
\hline       
\hspace{0.5cm} & $\chi^2 / d.o.f$               & $148.7 / 153$           & $ 49.9 / 45 $ \\
\hline       
\end{tabular}
\tablenotetext{a}{$10^{-2}$ counts sec$^{-1}$ in 0.3 -- 10 keV }\\
\tablenotetext{b}{Absorption corrected 0.5 -- 10 keV flux in $10^{-14}$\flux.}\\
\tablenotetext{c}{Absorption corrected 0.5 -- 10 keV luminosity in $10^{40}$\lum.}\\
\tablenotetext{d}{Hydrogen column density of the intrinsic absorber in $10^{23}$\colden.}\\
\tablenotetext{e}{Absorption corrected 2 -- 10 keV flux in $10^{-12}$\flux.} \\
\tablenotetext{f}{Absorption corrected 2 -- 10 keV luminosity in $10^{42}$\lum.} \\
\end{table}

\clearpage
\begin{table}[htbp]
\caption{Summary of PL and Bremss fits to the EPIC spectra 
of the lobes.} 
\label{table:lobe}
\begin{tabular}{lcccccc}
\hline\hline 
                               & \multicolumn{3}{c}{Northern Lobe}                                &   \multicolumn{3}{c}{Southern Lobe}   \\
\hline       
MOS Count Rate\tablenotemark{a}& \multicolumn{3}{c}{ $ 4.2  \pm 0.6 $ }                              &   \multicolumn{3}{c}{ $ 1.4 \pm 0.6 $ }   \\
PN  Count Rate\tablenotemark{a}& \multicolumn{3}{c}{ $ 17.5 \pm 2.2$ }                              &   \multicolumn{3}{c}{ $ 14.3 \pm 2.5 $ }   \\
\hline       
Model                          & PL                  & PL                   & Bremss              & PL                  & PL             & Bremss  \\
$\Gamma_{\rm X}$ or $kT$ (keV) & $2.2_{-0.5}^{+0.6}$ & $1.73$ (fix)         & $2.5_{-1.7}^{+6.5}$ & $1.7_{-0.6}^{+0.7}$ & $1.73$ (fix)   & $ 4.3_{-1.1}^{+\infty}$ \\
$F_{\rm X}$ ($10^{-13}$ \flux)\tablenotemark{b} & $4.6_{-1.5}^{+1.8}$ & $6.0_{-0.9}^{+1.0}$  & $4.4_{-1.0}^{+3.9}$ & $5.2_{-2.5}^{+3.1}$ & $5.0\pm1.5$    & $4.6_{-1.4}^{+2.7}$     \\
$L_{\rm X}$ ($10^{40}$ \lum)\tablenotemark{b}   & $8.3_{-2.6}^{+3.3}$ & $10.8 \pm 1.7$       & $8.1_{-1.9}^{+7.1}$ & $9.2_{-4.3}^{+5.7}$ & $9.0 \pm 2.6 $ & $8.3_{-2.5}^{+4.8}$     \\ 
\hline       
$\chi^2/{\rm d.o.f}$           &  $42.8 / 35 $       &  $45.2 / 36 $        & $48.4 / 35 $        & $10.6/13$           & $10.6/14$      &   $10.7/13$      \\ 
\hline       
\tablenotetext{a}{10$^{-3}$ counts sec$^{-1}$ in 0.5 -- 4.0 keV (MOS) and in 0.5 -- 6.0 keV (PN), except for 1.5 -- 1.7 keV (see text)}
\tablenotetext{b}{The best-fit model flux and luminosity in 0.7 -- 7 keV, evaluated after correcting the Galactic absorption. }
\end{tabular}
\end{table}

\clearpage
\begin{table}[htbp]
\caption{Physical parameters in the lobes of 3C 98.}
\label{table:param}
\begin{tabular}{llll}
\hline\hline 
                                  & Northern              & Southern              &    \\ 
\hline 
Size                              & $0.75\times1.25$      & $  0.67\times1.45$    & arcmin $\times$ arcmin \\
Volume				  & 3.3                   & 3.1                & $10^{69}$ cm$^{3}$   \\
$S_{\rm R}$\tablenotemark{a}      & $5.6$                 & $ 4.8$	          & Jy  \\
$\Gamma_{\rm R}$                      & \multicolumn{2}{c}{$1.73\pm0.01$}         &   \\
$B_{\rm me}$                      & $4.3$                 & $ 4.2$                & $\mu$G   \\
$S_{\rm 1keV}$                    & $8.5\pm1.4$           & $7.1\pm2.1$           & nJy   \\
\hline 
\ue                               & $ 6.1_{-0.9}^{+1.0} $ & $ 5.5\pm1.6 $         & $10^{-12}$~\enedens \\
\um                               & $ 1.2_{-0.1}^{+0.3} $ & $ 1.3_{-0.4}^{+0.6} $ & $10^{-13}$~\enedens \\
$B$                               & $ 1.7_{-0.1}^{+0.2} $ & $ 1.8_{-0.3}^{+0.4} $ & $\mu$G    \\
\hline 
\end{tabular}
\tablenotetext{}{Systematic uncertainties discussed in \S \ref{sec:emission} are not included.}\\ 
\tablenotetext{a}{Flux density at 1.4 GHz estimated from the best-fit radio PL spectrum.}\\
\end{table}

\clearpage

\clearpage

\clearpage

\clearpage

\clearpage

\clearpage


\begin{thebibliography}{}
\bibitem[]{3C219}
        Brunetti, G., Comastri, A., Setti, G., \&  Feretti, L.
        1999, \aap, 342, 57 
\bibitem[]{3C295}
        Brunetti, G., Cappi, M., Setti, G., Feretti, L., \& Harris, D.E.
        2001, \aap, 372, 755
\bibitem[]{lobe_thermal_2}
        Burch, S.F., 1979, \mnras, 186, 519 
\bibitem[]{3C219_2} 
	Comastri, A., Brunetti, G., Dallacasa, D., 
	Bondi, M., Pedani, M., \& Setti, G.,
        \mnras, 340, L52
\bibitem[]{3C66B_3C449}
	Croston, J.H., Birkinshaw, M., Hardcastle, M.J., \& Worrall, D. M., 
	2003, \mnras, 346, 1041
\bibitem[]{3C223}
	Croston, J.H., Birkinshaw, M., Hardcastle, M.J., \& Worrall, D. M., 
	2004, \mnras, 353, 879
\bibitem[]{FRII_study}
        Croston, J.H., Hardcastle, M.J., Harris D.E., Belsole, E., 
	Birkinshaw, M., \& Worrall, D. M., 
	2005, \apj, 626, 733
\bibitem[]{FR}
        Fanaroff, B.L., \& Riley, J.M.,
        1974, \mnras, 167, 31
\bibitem[]{fea95}
        Feigelson, E.D., Laurent-Muehleisen, S.A., Kollgaard, R.I., 
        \& Fomalont, E.B.,
        1995, \apjl, 449, L149 
\bibitem[]{M84}
        Finogurnov, \& Jones, 2001, \apj, 547, L107  
\bibitem[]{PSF}
        Ghizzardi, S. 2001,
	``In Flight Calibration of the PSF for the MOS1 and MOS2 Cameras''
	\xmm~Calibration Documentation (XMM-SOC-CAL-TN-0022)
\bibitem[]{jet-counter-jet}
        Giovannini, G.,  Cotton, W. D.,  Feretti, L.,
	 Lara, L., \& Venturi, T., 2001, \apj, 552, 508 
\bibitem[]{IR}
        Golombek, D., Miley, G.K., \& Neugebauer, G.,
	1988, \aj, 95, 26
\bibitem[]{Newton_PicA}
        Grandi, P., et al., 
	2003 \apj, 586, 123 
\bibitem[]{CMB_IC}
        Harris, D.E., and  Grindlay, J.E.,
        1979, \mnras, 188, 25
\bibitem[]{core_lobe_ratio}
        Hardcastle, M.J., Alexander,P., Pooley, G.G., \& Riley, J. M.
	1998 \mnras, 296, 445  
\bibitem[]{3CRR}
        Hardcastle, M.J., \& Worrall, D.M.,
	1999 \mnras, 309, 969
\bibitem[]{HBCetal}
        Hardcastle, M.J., Birkinshaw, M., Cameron, R.A.,
        Harris, D.E., Looney, L.W., \& Worrall, D.M.,
        2002, \apj, 581, 948
\bibitem[]{HS_summary}
        Hardcastle, M.J., Harris, D.E., Worrall D.M., \& Birlinshaw, M.,
	2004, \apj, 612, 729
\bibitem[]{IR_obscure}
        Heckman, T.M., O'Dea, C.P., Baum, S.A., \& Laurikainen, E.,
	1995, \apj, 428, 65 
\bibitem[]{3C452}
        Isobe, N., et al., 
        2002, \apjl, 581, L111
\bibitem[]{IsobeD}
        Isobe, N., 2002, 
        PhD thesis, University of Tokyo
\bibitem[]{kea95}
        Kaneda, H., 
        et al.          
        1995, \apjl, 453, L13
\bibitem[]{takaoka04}
	Kataoka, J., \& Stawarz, L., 
	\apj, 622, 797
\bibitem[]{CenA}
        Kraft, R.P., Vazquez, S.E., Forman W.R., Jones, C., \& Murray S.S., 
	2003, \apj, 592, 129  
\bibitem[]{1.4GHz_flux}
        Laing, R. A., \& Peacock, J. A.,
        1980, \mnras, 190, 903
\bibitem[]{SR_index}
        Laing, R. A., Riley, J. M., \& Longair, M.,
        1983, \mnras,  204, 151
\bibitem[]{VLA_immage}
        Leahy, J. P., et al., 
	1997, \mnras, 291, 20       
\bibitem[]{lobe_1.4GHz}
	Mackay, C.D., 
	1969, \mnras, 145, 31 
\bibitem[]{ellitical}
        Matsushita, K., Ohashi, T., \&  Makishima, K.,
        2000, \pasj, 52, 685
\bibitem[]{HydraA}
	McNamara, B.R., et al.,
	2000, \apjl, 534, L135
\bibitem[]{Bme}
        Miley, G., 1980, \araa, 18, 165 
\bibitem[]{RG_ASCA}
        Sambruna, R.M., Eracleous, M. \& Mushotzky, R.F.,
	1999, \apj, 526, 60
\bibitem[]{redshift}
        Schimidt, M., 1965, \apj, 141, 1 
\bibitem[]{lobe_thermal_3}
        Spangler, S.R., \& Sakurai, T.,
	1985, \apj, 297, 84  
\bibitem[]{tea98}
        Tashiro, M., et al, 
	1998, \apj, 499, 713
\bibitem[]{tea01}
        Tashiro, M., Makishima, K., Iyomoto, N.
        Isobe, N., \&  Kaneda, H. 2001, \apjl, 546, L19 
\bibitem[]{host_gal}
        Zirbel, E.L., 1996, \apj, 473, 713
\end{thebibliography}
\end{document}